\documentclass[AMA,twocolumn]{USG}
\usepackage{lettersp}
\usepackage[numbers,super]{NJDnatbib}

\setcitestyle{numbers,square}
\makeatletter\def\NAT@sort{\@ne}\def\NAT@cmprs{\@ne}\makeatother %

\newif\ifsubmission
\ifdefined\SubmissionBuild \submissiontrue \else \submissionfalse \fi

\usepackage[no-math]{fontspec}
\AtBeginDocument{\tracinglostchars=3\relax}

\usepackage{totcount}
\usepackage{lastpage}
\regtotcounter{table}
\regtotcounter{figure}
\ifsubmission
  \usepackage[switch]{lineno}
\fi

\makeatletter
\renewcommand{\@maketitle}{%
  {%
    \hsize=\textwidth\parindent=0\@p@t
    \thispagestyle{titlepage}%
    \let\footnote\thanks
    \@@articletype
    \vskip8.6mm%
    \ifx\@title\empty\else{\@title\par}\fi
    \removelastskip\vskip8\@p@t
    {\artauthors\par}%
    \removelastskip\vskip27.25\@p@t
    \printabstract
    \removelastskip\vskip\adjtitleskip
  }%
}
\def\@@articletype{%
  \rule[-4\@p@t]{4\@p@t}{15\@p@t}%
  {\arttypefont\bf\uppercase{\letterspace to 1.15\naturalwidth{\@articletype}}}%
}
\def\oddfoot@titlepage@info{\oddhead@headings@info}
\def\evenhead@headings@info{\head@headings@def{\evenpage@headings@def\hfill}}
\def\thepagedef{{\pagenumfont\thepage\ of \pageref*{LastPage}}}
\def\CAT@addrsep{\unskip\ \ $|$\ \ }
\renewcommand{\@@address}[2][]{%
  \g@addto@macro\jmkaddress{%
    \hsize=\textwidth
    \def\baselinestretch{1}%
    \stepcounter{affn}%
    \xdef\@currentlabel{\theaffn}%
    \jmkLabel{#1}%
    \addressfont
    \ifnum\value{affn}>\@ne\CAT@addrsep\fi
    \if\relax\detokenize{#1}\relax\else\textsuperscript{\smash{\theaffn}}\fi
    #2%
  }%
}
\g@addto@macro\maketitle{\gdef\authorsep{}}
\historydates{}
\def\bwfiginfofont{\rmfamily\fontsize{10\@p@t}{10\@p@t}\selectfont}

\def\CAT@wide{wide}
\newcommand{\DefineFigure}[4]{%
  \@ifundefined{CAT@fig@body@#1}{}{%
    \PackageError{main}{Figure `#1' defined twice}{Each key is defined once in main.tex.}}%
  \expandafter\def\csname CAT@fig@span@#1\endcsname{#2}%
  \expandafter\long\expandafter\def\csname CAT@fig@body@#1\endcsname{#3}%
  \expandafter\long\expandafter\def\csname CAT@fig@legend@#1\endcsname{#4}%
}
\newcommand{\DefineTable}[4]{%
  \@ifundefined{CAT@tab@body@#1}{}{%
    \PackageError{main}{Table `#1' defined twice}{Each key is defined once in main.tex.}}%
  \expandafter\def\csname CAT@tab@span@#1\endcsname{#2}%
  \expandafter\long\expandafter\def\csname CAT@tab@body@#1\endcsname{#3}%
  \expandafter\long\expandafter\def\csname CAT@tab@legend@#1\endcsname{#4}%
}
\DeclareRobustCommand{\FigureLegend}[1]{\@nameuse{CAT@fig@legend@#1}}
\DeclareRobustCommand{\TableLegend}[1]{\@nameuse{CAT@tab@legend@#1}}
\def\CAT@checkdefined#1#2{%
  \@ifundefined{CAT@#1@body@#2}{%
    \PackageError{main}{Float `#2' is not defined}{Define it with \DefineFigure/\DefineTable before it is placed.}}{}}
\let\CAT@figqueue\@empty
\let\CAT@tabqueue\@empty
\newcommand{\CAT@aboutmark}[1]{%
  \par\medskip{\centering[#1 about here]\par}\medskip}
\newcommand{\placefigure}[1]{%
  \CAT@checkdefined{fig}{#1}%
  \ifsubmission
    \refstepcounter{figure}\label{fig:#1}%
    \g@addto@macro\CAT@figqueue{\CAT@shipfigure{#1}}%
    \CAT@aboutmark{Figure~\thefigure}%
  \else
    \edef\@tempa{\csname CAT@fig@span@#1\endcsname}%
    \ifx\@tempa\CAT@wide\def\@tempb{figure*}\else\def\@tempb{figure}\fi
    \begin{\@tempb}[tbp]%
      \centering\@nameuse{CAT@fig@body@#1}%
      \caption{\FigureLegend{#1}}\label{fig:#1}%
    \end{\@tempb}%
  \fi
}
\newcommand{\placetable}[1]{%
  \CAT@checkdefined{tab}{#1}%
  \ifsubmission
    \refstepcounter{table}\label{tab:#1}%
    \g@addto@macro\CAT@tabqueue{\CAT@shiptable{#1}}%
    \CAT@aboutmark{Table~\thetable}%
  \else
    \edef\@tempa{\csname CAT@tab@span@#1\endcsname}%
    \ifx\@tempa\CAT@wide\def\@tempb{table*}\else\def\@tempb{table}\fi
    \begin{\@tempb}[tbp]%
      \caption{\TableLegend{#1}}\label{tab:#1}%
      \centering\@nameuse{CAT@tab@body@#1}%
    \end{\@tempb}%
  \fi
}
\newcommand{\CAT@shiptable}[1]{%
  \clearpage
  {\raggedright\noindent\textbf{Table~\ref*{tab:#1}.}\ \TableLegend{#1}\par}%
  \bigskip
  {\centering\@nameuse{CAT@tab@body@#1}\par}%
}
\newcommand{\CAT@legendentry}[1]{%
  \par\noindent\textbf{Figure~\ref*{fig:#1}.}\ \FigureLegend{#1}\par\medskip}
\newcommand{\CAT@shipfigure}[1]{%
  \clearpage
  \vspace*{\fill}%
  {\centering\@nameuse{CAT@fig@body@#1}\par}%
  \bigskip
  {\centering\textbf{Figure~\ref*{fig:#1}}\par}%
  \vspace*{\fill}%
}
\newcommand{\printfloats}{%
  \ifsubmission
    \onecolumn
    \nolinenumbers
    \CAT@tabqueue
    \ifx\CAT@figqueue\@empty\else
      \clearpage
      \linenumbers
      \section*{Figure Legends}%
      \def\CAT@shipfigure##1{\CAT@legendentry{##1}}%
      \CAT@figqueue
      \nolinenumbers
      \let\CAT@shipfigure\CAT@shipfigure@page
      \CAT@figqueue
    \fi
  \fi
}
\AtBeginDocument{\let\CAT@shipfigure@page\CAT@shipfigure}

\newcommand{\FloatGraphic}[1]{%
  \sbox\z@{#1}%
  \ifdim\wd\z@>\linewidth\resizebox{\linewidth}{!}{\usebox\z@}\else\usebox\z@\fi}

\newcommand{\printreferences}{%
  \bibliography{references}}
\makeatother

\articletype{ORIGINAL ARTICLE}

\title{A Frame-Level View and Image-Quality Annotation Tool for Fetal Echocardiography: Comparison with Spreadsheet Annotation}
\titlemark{Frame-level view and quality annotation in fetal echocardiography}
\authormark{Azzam \textsc{et al.}}

\author[a1]{Mohamed Azzam}
\author[a1]{Ziyang Xu}
\author[a2]{Quinlan Couch}
\author[a2]{Zoe Keese}
\author[a2]{Andrea Cadlo}
\author[a2]{Christine Sedze-DeWolf}
\author[a2]{Micaiah Uhrmacher}
\author[a2]{Madelyn Graff}
\author[a2]{Catherine Fox}
\author[a2]{Lauren Vatter}
\author[a2]{Paige Jeffryes}
\author[a3,a4,a5]{Jason Christensen}
\author[a2]{Neil Hamill}
\author[a6]{Shibiao Wan}
\author[a3,a4,a5]{Ling Li}
\author[a1]{Jieqiong Wang}

\address[a1]{\orgdiv{Department of Neurological Sciences, }\orgname{University of Nebraska Medical Center, }\orgaddress{\city{Omaha, }\state{NE 68198, }\country{USA}}}
\address[a2]{\orgdiv{Department of Obstetrics and Gynecology, }\orgname{University of Nebraska Medical Center, }\orgaddress{\city{Omaha, }\state{NE 68198, }\country{USA}}}
\address[a3]{\orgdiv{Department of Pediatrics, }\orgname{University of Nebraska Medical Center, }\orgaddress{\city{Omaha, }\state{NE 68198, }\country{USA}}}
\address[a4]{\orgname{Criss Heart Center, Children's Nebraska, }\orgaddress{\city{Omaha, }\state{NE 68114, }\country{USA}}}
\address[a5]{\orgname{Child Health Research Institute, }\orgaddress{\city{Omaha, }\state{NE 68198, }\country{USA}}}
\address[a6]{\orgdiv{Department of Genetics, Cell Biology and Anatomy, }\orgname{University of Nebraska Medical Center, }\orgaddress{\city{Omaha, }\state{NE 68198, }\country{USA}}}

\corres{Jieqiong Wang (\mbox{jiwang@unmc.edu})}

\keywords{fetal echocardiography | congenital heart disease | image annotation | standard views | image quality | artificial intelligence}

\abstract[ABSTRACT]{%
  \textbf{Objective:} To develop the Cardiologist Annotation Tool (CAT)
  for frame-level cardiac view and image-quality annotation in fetal
  echocardiography and compare it with spreadsheet-based annotation.\par

  \textbf{Methods:} CAT provides cine playback, frame navigation, and
  keyboard-based labeling without manual frame-range entry. Experts %
  annotated 25 studies using both methods, and spreadsheet annotations
  were audited for predefined data-entry errors. In a randomized
  crossover study, six medical students annotated two normal studies,
  one with each method, against an expert reference.\par

  \textbf{Results:} Among 1,746 spreadsheet-annotated files, 6.2\%
  contained major or critical data-entry errors, more frequently in
  congenital heart disease than normal studies (10.9\% vs 3.4\%).
  CAT yielded more annotated view segments per clip (3.6 vs 1.5).
  Rates of labeled and correctly labeled frames per minute were 1.77
  and 1.68 times those with spreadsheets, respectively
  (both \textit{P} = 0.031), with gains in all six students.
  No detectable difference in view-label accuracy was observed.\par

  \textbf{Conclusion:} CAT eliminates manual frame-range entry and %
  enables finer-grained, structured annotation of fetal
  echocardiography, providing a practical approach to preparing
  datasets for AI research.%
}

\begin{document}

\maketitle
\ifsubmission\linenumbers\fi

\begin{boxwithhead}{Key points}{%
  \textbf{What's already known about this topic?}
  \begin{itemize}
    \item{} Prenatal heart disease detection models need fetal echocardiography frames
      labeled by experts with the standard view, ideally with image quality.
  \end{itemize}
  \textbf{What does this study add?}
  \begin{itemize}
    \item{} Dedicated annotation software labels view and image quality per frame on DICOM cine,
      preventing spreadsheet data-entry errors by construction.
    \item{} Major or critical errors affected 6.2\% of expert spreadsheet files; a student
      crossover estimated efficiency and accuracy.
  \end{itemize}
}
\end{boxwithhead}

\section{Introduction}\label{sec:introduction}
Congenital heart disease (CHD) affects approximately 9 per 1,000
live births~\cite{vanderLinde2011Birth,Liu2019Global} and is a
leading cause of infant morbidity and mortality~\cite{Carvalho2023ISUOG,GBD2020Burden}.
Prenatal detection
supports delivery planning at centers with neonatal cardiac
care~\cite{Holland2015Prenatal}, but detection rates vary widely.
For critical CHD, reported rates range from 13\% to 87\% across
international screening programs~\cite{Bakker2019Prenatal}
and from 12\% to 53\% across US states among infants undergoing
cardiac surgery~\cite{Quartermain2015Variation}. Including
outflow-tract and three-vessel views alongside the four-chamber
view improves detection~\cite{Donofrio2014Diagnosis,Carvalho2023ISUOG}.
Nevertheless, lesions requiring these additional views are
less consistently detected, and performance depends on examiner
training~\cite{Quartermain2015Variation,Donofrio2014Diagnosis}.

Software has been developed to support prenatal diagnosis
from ultrasound findings~\cite{Abdelnaby2017SonoCAD}, while
artificial intelligence (AI) offers potential support for
CHD detection from ultrasound images. Models that identify
standard planes in freehand fetal ultrasound~\cite{Baumgartner2017SonoNet}
and detect CHD from cardiac
views~\cite{Arnaout2021Ensemble} rely on identifying relevant
frames within ultrasound studies. Developing and evaluating
these models requires expert-labeled cardiac views, ideally
accompanied by image-quality annotations to identify
diagnostically useful frames~\cite{Arnaout2021Ensemble,Fiorentino2023Review}.

However, fetal echocardiography datasets with both frame-level
view and quality annotations remain limited
(Section~\ref{sec:related-work}). Producing these labels is
labor-intensive and requires substantial expert time~\cite{Willemink2020Preparing,Zhang2024Advances}.
An efficient workflow should allow fetal cardiologists to
focus on image interpretation rather than manually recording
frame ranges and labels. In the spreadsheet workflow evaluated
here, experts review cine in a separate viewer and type
annotations into spreadsheets. Without checks against file
length and existing labels, overlapping, missing, or
out-of-range entries may go unnoticed.

\DefineFigure{figure1_overview}{wide}
{%
  \FloatGraphic{\includegraphics{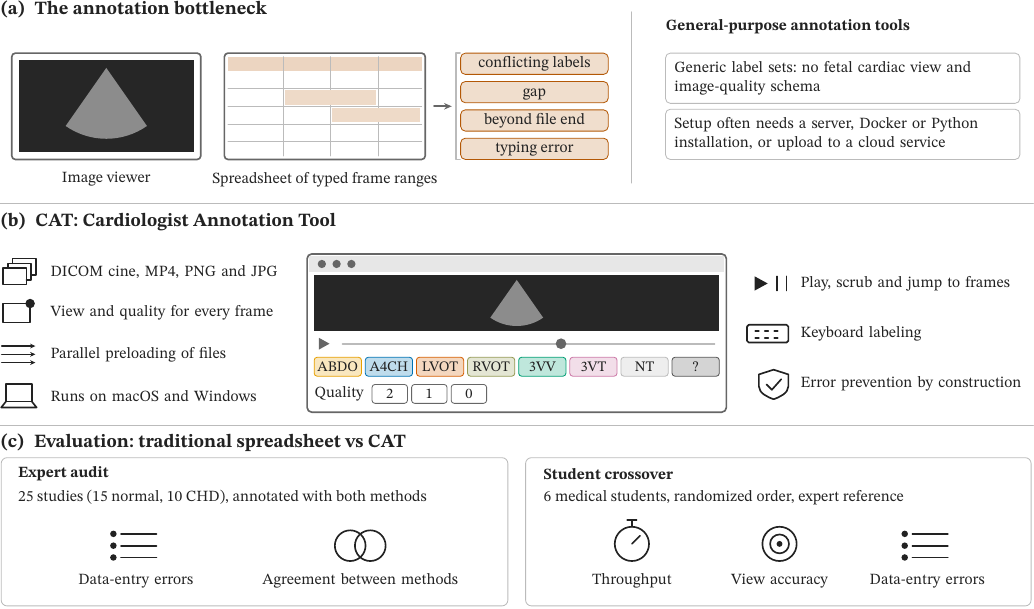}}%
}
{%
  Study overview. \textbf{(a)}~In traditional annotation, frame ranges are typed into a
  spreadsheet beside an image viewer, which allows structural data-entry errors; general-purpose
  annotation tools offer generic label sets and need setup. \textbf{(b)}~The Cardiologist
  Annotation Tool (CAT) records the view and image quality of every frame directly on DICOM cine and common video and image formats. \textbf{(c)}~CAT was compared with the
  traditional spreadsheet in an expert audit and a randomized crossover study of medical
  students. CHD, congenital heart
  disease; view abbreviations and quality levels as in
  Figure~\ref{fig:figure2_cat_software}.%
}
To address these challenges, we developed the Cardiologist
Annotation Tool (CAT) in close collaboration with fetal
cardiologists to make efficient use of their limited annotation
time. CAT reads DICOM cine directly, preloads neighboring files,
and supports keyboard-based labeling of cardiac views and
image quality (Figure~\ref{fig:figure1_overview}).
Annotations are stored as change points, ensuring that
each frame receives a single view and quality label while
preventing structural data-entry errors by design.
\placefigure{figure1_overview}

This study presents three contributions:
\begin{enumerate}
  \item A dedicated fetal echocardiography annotation tool that
  records cardiac views and image quality frame by frame directly
  on DICOM cine and common video and image formats, while preventing structural data-entry errors
  by design. To our knowledge, no published tool combines these features.

  \item An expert audit of 25 fetal echocardiography studies
  (15 normal and 10 CHD) comparing CAT with spreadsheet
  annotation. Using a predefined error codebook and
  validated detectors, we quantified data-entry errors,
  examined their association with CHD, and assessed
  annotation agreement and granularity.

  \item A randomized crossover study involving six medical
  students to compare annotation throughput and view
  accuracy against an expert reference. Frame-based
  throughput was higher with CAT for all six students,
  without a detectable difference in view accuracy.
\end{enumerate}

\section{Related Work}\label{sec:related-work}

\subsection{Annotation practices in fetal ultrasound}

Published fetal cardiac datasets illustrate different approaches
to expert labeling. Arnaout et al.\ assigned every frame to one
of five screening views or a non-target category and used only
diagnostically adequate views to train their view classifier~\cite{Arnaout2021Ensemble}.
For HeartAssist, one expert
classified 65,324 images from 2,985 fetuses into ten cardiac
views~\cite{Kim2025HeartAssist}. Taks{\o}e-Vester et al.\
used LabelMe to annotate still images of eight cardiac planes
and assign structure-level quality scores~\cite{TaksoeeVester2024Quality},
while Nurmaini et al.\
outlined cardiac structures in frames extracted from ultrasound
videos~\cite{Nurmaini2021Deep}. Other fetal ultrasound datasets
used customized CVAT or in-house interfaces to annotate
images~\cite{Prabakaran2023FPUS23,Pu2026FUSEP,
BurgosArtizzu2020Evaluation}. These studies demonstrate
variation in annotation granularity and how image quality
is assessed.

\subsection{Existing annotation tools}

General-purpose tools such as LabelMe, VIA, CVAT, and Label
Studio support image labeling, video-frame annotation, or
temporal segments~\cite{Russell2008LabelMe,LabelmeGitHub2026,Dutta2019VIA,
CVATDocs2026,LabelStudioDocs2026,Tkachenko2020LabelStudio}. Their flexibility
allows project-specific label schemas, but fetal cardiac
views and image-quality grades require configuration.

Ultrasound-specific tools provide more tailored capabilities.
Annotation Web supports ultrasound-video annotation, including
sequence-level image-quality fields, and has been adapted for
regional quality scoring in adult echocardiography~\cite{Smistad2021AnnotationWeb,VanDeVyver2025Regional}.
SlicerUltrasound processes DICOM ultrasound cine and can attach
classification labels to entire loops~\cite{SlicerUltrasoundGitHub2026}.
Most directly relevant,
an Oxford fetal heart tool labels each video frame with one of
three cardiac views and a visibility state. It works with
screen-captured video rather than native DICOM cine and does
not record a frame-level quality grade~\cite{Bridge2021HeartAnnotation,Bridge2017Thesis}.
Across these tools, support for native DICOM cine, frame-level
fetal cardiac view and image-quality labeling, and prevention
of structural data-entry errors remains distributed across
different platforms. None of the reviewed tools reported an
evaluation against another annotation method. To address these
gaps, CAT was designed to combine these features in a portable
desktop application for macOS and Windows, without requiring
package installation or annotation configuration. We evaluated
CAT against traditional spreadsheet annotation in complementary
expert and student studies.

\subsection{Automated and weakly supervised annotation}

Other approaches reduce the need for manual frame-level
annotation. SonoNet derived standard-plane labels from
freeze frames saved during routine scanning~\cite{Baumgartner2017SonoNet}.
The iFIND and PULSE projects
used freeze events, on-screen text, and machine settings
to recover information from recorded scans~\cite{Venturini2025Whole,Sharma2021Knowledge,
Drukker2021Transforming}. In adult echocardiography,
stored DICOM view tags have also provided labels~\cite{UlloaCerna2021Deep}.
Another study used AI to extract video clips of the standard cardiac views from full fetal
scans for remote second review by experts~\cite{Day2025Video}. These approaches can
accelerate dataset preparation, but stored labels reflect what sonographers chose to save,
and standard-plane detectors trained on normal hearts may miss frames that show structural
defects~\cite{Tan2020Automated}. Expert-reviewed frame-level labels therefore remain
valuable for evaluating automated view and quality predictions.

\section{Methods}\label{sec:methods}
We describe the tool and then its two evaluations, an audit of expert annotations and a
crossover study in medical students (Table~\ref{tab:table2_study_design}).

\DefineTable{table2_study_design}{wide}
{%
  \begin{minipage}{\textwidth}
  \fontsize{8}{10}\selectfont
  \begin{tabularx}{\textwidth}{@{}>{\raggedright\arraybackslash}p{27mm}
                                  >{\raggedright\arraybackslash}X
                                  >{\raggedright\arraybackslash}X@{}}
    \toprule
    & \textbf{Expert audit} & \textbf{Student crossover study} \\
    \midrule
    Question
      & How often spreadsheet labels contain data-entry errors, and how spreadsheet and
        CAT labels agree
      & Whether CAT changes annotation throughput and view accuracy \\
    \addlinespace
    Annotators
      & Four pediatric and fetal cardiologists (mean 17.5 years of experience): two in the
        traditional pass and three in the CAT pass; one took part in both
      & Six medical students, trained by an expert on one example scan \\
    \addlinespace
    Cases
      & 25 studies (15 normal, 10 CHD); 1,746 files annotated with both methods
      & Two normal scans (27 and 29 valid files), with invalid decoy files to be left
        unannotated \\
    \addlinespace
    Design
      & Original DICOM cine, annotated first on the experts' own spreadsheet and 3 to 5
        months later with CAT
      & Randomized crossover on simplified video and image copies: three students used
        the spreadsheet template first and three used CAT first, with one day or less
        between sessions \\
    \addlinespace
    Comparison standard
      & %
      CAT serves as a comparator, and disagreements are unadjudicated
      & Expert reference, created with CAT and revised twice by experts \\
    \addlinespace
    Annotation time
      & Not recorded
      & Recorded by each student in a template spreadsheet; checked against CAT's event
        log for CAT sessions \\
    \addlinespace
    Outcomes
      & Data-entry errors; agreement between the methods on the view, image quality and
        view boundaries
      & Annotation time; reference segments, labeled frames and correct frames per
        minute; view accuracy; per-view recall; data-entry errors \\
    \bottomrule
  \end{tabularx}
  \end{minipage}%
}
{%
  Design of the expert audit and the student crossover study%
}
\placetable{table2_study_design}

\subsection{Development of the annotation tool}
CAT was developed in close collaboration with fetal cardiologists to
maximize the efficiency of their limited annotation time. The software
was implemented in Python with a Tkinter-based graphical user interface
and is distributed as a bundled executable for macOS and Windows that runs
locally on the annotator's computer without installing Python or other packages. The initial version was
developed in one month roughly, followed by two months
of testing and iterative refinement based on feedback from fetal
cardiologists. During this period, additional features were incorporated
to address their annotation needs and improve usability. The finalized
version was used consistently in both experimental studies.

\subsection{Software and annotation workflow}
CAT opens a study folder, validates its files,
and reads DICOM cine files directly, as well
as MPEG-4 (MP4) video and image files (Figure~\ref{fig:figure2_cat_software}). While one file is
displayed, neighboring files are decoded in background threads (preloading). The annotator
plays or scrubs through the cine and labels frames with single keys; changes are saved
automatically, and labels are exported as comma-separated values (CSV) files.

\DefineFigure{figure2_cat_software}{wide}
{%
  \FloatGraphic{\includegraphics{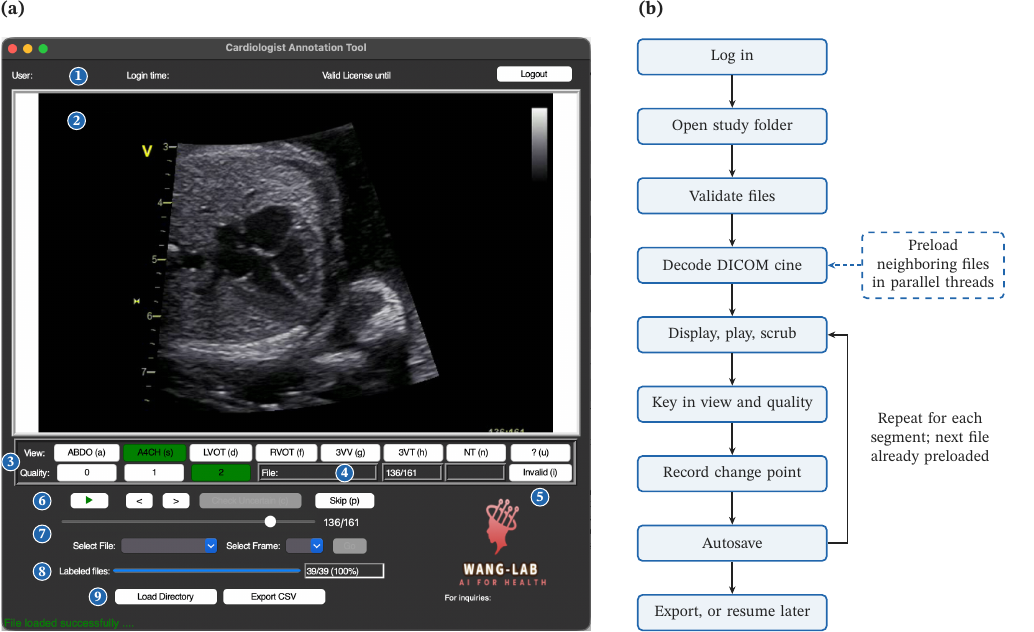}}%
}
{%
  The CAT software.
  \textbf{(a)}~Main window during annotation of a fetal
  echocardiography study where the displayed frame
  is labeled A4CH, quality 2. Callouts: 1, session bar; 2, cine frame; 3, view keys (top)
  and quality keys (bottom), keyboard keys in parentheses; 4, file and frame
  number; 5, mark file invalid; 6, playback, Check Uncertain and Skip (leave the
  file for later); 7, frame slider and jump to a file or frame; 8, progress
  through the study; 9, open a study folder and export the annotations.
  \textbf{(b)}~Annotation workflow. Dashed blocks denote background step. Each label is stored
  as a change point and holds until the next label in the file.
  Views: abdominal (ABDO), apical four-chamber (A4CH), left (LVOT) and right (RVOT) ventricular
  outflow tracts, three-vessel (3VV), three-vessel and trachea (3VT), non-target (NT),
   uncertain (?). Quality: 2 for clearly visible,  1 for acceptable with some noise, and  0 for not adequately visible.%
}
\placefigure{figure2_cat_software}

\subsection{Annotation model}
Each frame receives one of eight view labels: the abdominal (situs) view (ABDO), apical
four-chamber view (A4CH), left ventricular outflow tract (LVOT), right ventricular outflow
tract (RVOT), three-vessel view (3VV), three-vessel and trachea view (3VT), non-target (NT) for
a frame that shows none of these six target views, and ``?'' (uncertain) for frames to
recheck. No frame was labeled ``?'' in either study. The target views correspond to the views
of the fetal cardiac screening examination~\cite{Carvalho2023ISUOG}; the aortic and ductal arch views are not
among them, so such frames were labeled NT. Target-view frames also receive a quality score on
a three-level ordinal scale, judged on the visibility of the cardiac structures that define the
view: 2, clearly visible without occlusion or noise; 1, some noise or not fully clear, but
acceptable; 0, not adequately visible. Quality has no meaning for NT frames and was ignored for
them in every analysis.

The annotator records a label only where it changes, and the label holds until the next change
point in the same file or to the end of the file, so the segments tile each file without gaps
or overlaps. When a file is first displayed, CAT records NT with quality 2 for its first frame.

\subsection{Error prevention by design}
Frame ranges typed into a spreadsheet beside a separate viewer can overlap under different
views, leave frames uncovered, run beyond the end of a file, or be unreadable. CAT prevents
these structural data-entry errors by construction (Table~\ref{tab:table3_error_prevention}):
each frame has one label, ranges derive from change points, frame numbers come from the
display, labels from fixed keys, and file names from disk. %

\DefineTable{table3_error_prevention}{wide}
{%
  \begin{minipage}{\textwidth}
  \fontsize{8}{10}\selectfont
  \begin{tabularx}{\textwidth}{@{}>{\raggedright\arraybackslash}p{43mm}
                                  >{\raggedright\arraybackslash}p{19mm}
                                  >{\raggedleft\arraybackslash}p{20mm}
                                  >{\raggedleft\arraybackslash}p{14mm}
                                  @{\hspace{4mm}}>{\raggedright\arraybackslash}X@{}}
    \toprule
    & & \multicolumn{2}{c}{\textbf{Findings in traditional spreadsheets}} & \\
    \cmidrule(lr){3-4}
    \textbf{Error class}
      & \textbf{Severity}
      & \textbf{Experts (files affected)}
      & \textbf{Students}
      & \textbf{How CAT prevents it by design} \\
    \midrule
    Conflicting labels on the same frames
      & major--critical & 49 (35) & 3
      & One label per frame: a label holds from its change point to the next \\
    Overlap at a shared endpoint
      & minor & 39 (36) & 6
      & Segment ranges are derived from change points, so adjacent segments cannot share a frame \\
    Duplicated range, same label
      & minor & 0 (0) & 1
      & Consecutive identical labels are merged \\
    Range beyond the file end
      & major--critical & 15 (14) & 0
      & Frame numbers are taken from the displayed frame; the last segment ends at the file's frame count \\
    Invalid range (reversed, or starting before the first frame)
      & major & 0 (0) & 1
      & Ranges are computed, not typed \\
    Unlabeled frames (gaps)
      & major & 65 (60) & 5
      & Segments tile the whole file \\
    File row with no labels
      & major & 8 (8) & 0
      & The first frame is labeled as soon as the file is displayed \\
    Unreadable or ambiguous entry
      & major--critical & 7 (7) & 1
      & Labels come from fixed buttons and keys \\
    Formatting irregularity
      & minor & 60 (45) & 10
      & Labels are saved in a fixed, structured format \\
    Row identity (duplicate or missing file name, omitted file)
      & major--critical & 0 (0) & 0
      & File names are read from disk \\
    \midrule
    Any provable major or critical error
      & major--critical & 138 (109) & 10 & \\
    \bottomrule
  \end{tabularx}
  \end{minipage}%
}
{%
  Data-entry error classes in traditional spreadsheet annotation and how CAT
  prevents each by design%
}
\placetable{table3_error_prevention}

\subsection{Studies}
\textbf{Expert audit.} Pediatric and fetal cardiologists annotated 25 fetal
echocardiography studies, selected at random within two groups: 15 normal
studies and 10 with CHD. In the traditional pass, two experts reviewed the original DICOM files in
an image viewer and typed frame ranges into a spreadsheet of their own design, with one row per
file and one column per view and quality level. After an interval of 3 to 5 months,
three experts annotated the same files with CAT where one expert took part in both passes. The four experts had 17.5 years of experience on average. In this setting, annotation time was not
measured and there was no reference standard to adjudicate disagreements between both methods. The analysis set comprised 1,746 files (247,477 frames) present in both sources.

\textbf{Student crossover study.} A total of six medical students each annotated two
normal scans, one with each method, in a randomized two-period crossover:
three students were randomly selected for Group A and the remaining three formed Group B. Group~A (student1--student3) used the traditional spreadsheet first and
Group~B (student4--student6) used CAT first. The students were not told which group they were
in or whether other students annotated the same scans, and each annotated separately from the
others. Beforehand, the students were trained by an expert on one example scan similar to the two study scans. Each scan included invalid decoy files, which the students were instructed to leave unannotated
(Table~\ref{tab:table2_study_design}). Each student received renamed, shuffled video and image
copies of the scans and recorded the start and end of each session in a template spreadsheet; the
spreadsheet template for traditional annotation restricted view and quality to the permitted
values. The students' labels were scored against the expert reference, which was created with CAT and
revised twice by experts. %

\subsection{Outcomes}
\textbf{Data-entry errors.} Before any error was counted, we froze an error codebook
of 21 codes in six families: formatting, unreadable or
ambiguous entries, range validity, overlaps, coverage gaps and row identity. Each code has a
certainty tier, provable (wrong by the spreadsheet's own logic or by the file length) or requiring
the images, and a severity graded by its effect on a
training dataset (critical: a wrong or contradictory view label, or labels on the wrong file;
major: labels lost or quality off by two levels; Table~\ref{tab:table3_error_prevention}).
The main measure was the share of files with a
provable finding of major or critical severity.

\textbf{Agreement between methods.} On frames with exactly one traditional view, we computed
view agreement, Cohen's $\kappa$ and Gwet's first-order agreement coefficient (AC1)~\cite{Cohen1960Coefficient,Gwet2008Computing}, and
on frames with the same target view in both methods, linearly weighted $\kappa$ for quality~\cite{Cohen1968Weighted}.
Boundary agreement was the share of view-change boundaries of one method that had a boundary
of the other method within $\pm k$ frames, for the range of tolerances $k$ shown in Figure~\ref{fig:figure4_expert_agreement}c. Each traditional frame was classified as
agreeing with CAT, differing from CAT, unlabeled, or conflicting. Because annotators and method
changed together, this represents a method-plus-rater agreement.

\textbf{Student measures.} Outcomes included self-reported annotation
time, reference view segments per minute, labeled and correctly labeled
frames per minute, view accuracy, per-view recall, and data-entry errors
identified using an adapted codebook. Reference segments per minute
was calculated using the fixed number of expert-defined segments,
independent of how students divided the clips. Correctly labeled frames
were those matching the expert reference. View accuracy was the
proportion of reference frames correctly labeled, with unlabeled or
conflicting frames counted as incorrect. Predefined flags identified
unreliable time records, protocol deviations (including annotation of
at least half the decoys), and outliers (robust $|z| > 3.5$).
Flagged observations were retained in the primary analysis.

\subsection{Statistical analysis}
In the expert audit, per-study error rates were assigned Wilson
95\% confidence intervals (CIs)~\cite{Wilson1927Probable}.
Pooled error rates and agreement measures were assigned CIs using
2,000 study-level bootstrap resamples stratified by CHD
status~\cite{Efron1993Introduction}. Error rates in normal and CHD
studies were compared using a permutation test (100,000 permutations).
A logistic generalized estimating equation (GEE)
model~\cite{Liang1986Longitudinal} assessed associations between
major or critical data-entry errors and CHD status, log frame count,
file type, and file position, accounting for clustering within studies.

In the student study, paired CAT-to-spreadsheet ratios were calculated
for annotation time and throughput measures, while paired differences
were used for view accuracy. Ratios were analyzed on the log scale.
We used exact two-sided paired permutation tests~\cite{Ernst2004Permutation}
and Hodges--Lehmann estimates with exact CIs (96.9\% for six
pairs)~\cite{Hodges1963Estimates,Hollander2013Nonparametric}.
\textit{P} values were not adjusted for multiplicity.
A separate randomized crossover analysis assessed the method effect
while accounting for the combined influence of scan and annotation
period. Sensitivity analyses included exclusion of the student with
unreliable timing, and CAT active time derived from its event logs. All analyses were conducted in Python.

\section{Results}\label{sec:results}

\subsection{Data-entry errors in the traditional spreadsheets}
Of the 1,746 files analyzed in both sources, 109 (6.2\%,
95\% CI 4.4--8.4) carried at least one provable data-entry error of
major or critical severity in the traditional spreadsheet
(Figure~\ref{fig:figure3_expert_errors}a, Table~\ref{tab:table3_error_prevention}). Such
errors occurred in 23 of the 25 studies, with per-study rates of up
to 25.0\% (Figure~\ref{fig:figure3_expert_errors}b).

\DefineFigure{figure3_expert_errors}{wide}
{%
  \FloatGraphic{\includegraphics{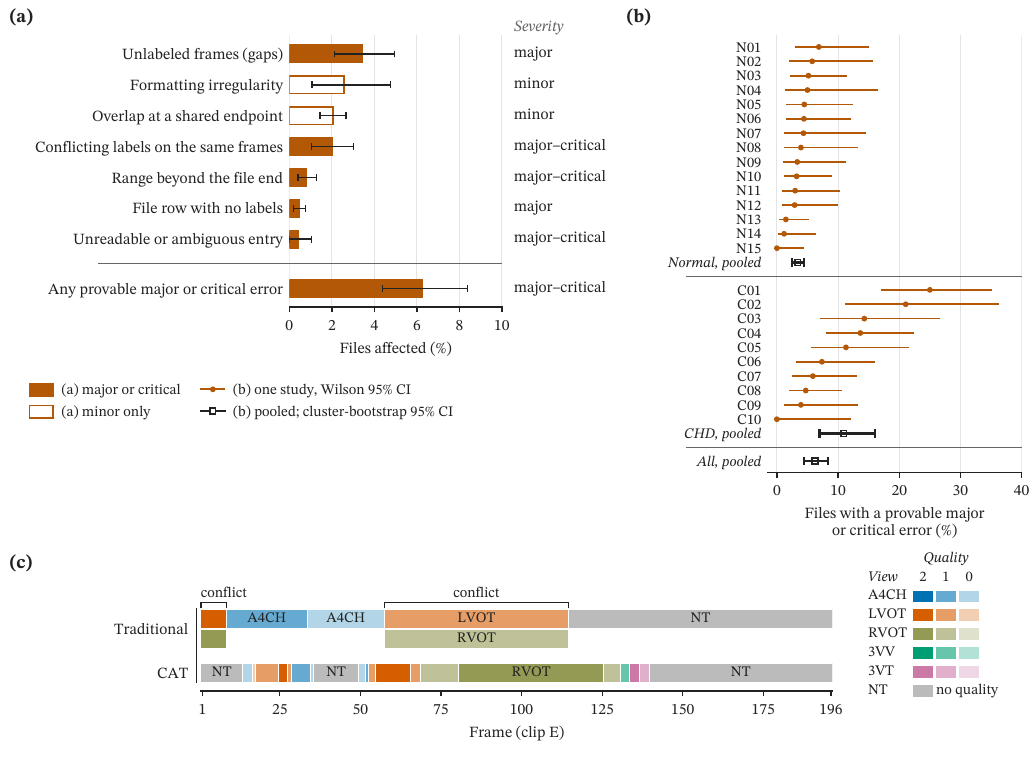}}%
}
{%
  Data-entry errors in the traditional spreadsheet annotations of the expert
  audit (1,746 files, 25 studies). CAT stores one label per change point and so prevents
  these errors by construction.
  \textbf{(a)}~Files with at least one error of each class, with each class's
  severity range.
  \textbf{(b)}~Files with a major or critical error, per study and
  per group (file-weighted).
  \textbf{(c)}~Clip~E, illustrating an example of view annotation in both methods: two
  traditional ranges give different views to the same frames; the overlapping range is
  drawn beneath (conflicting labels). CAT annotations instead prevent this error and indicates more fine-grained annotation.
  Intervals: 95\% confidence intervals (CIs), Wilson per study, otherwise
  study-level cluster bootstrap. View and quality abbreviations as in
  Figure~\ref{fig:figure2_cat_software}.%
}
\placefigure{figure3_expert_errors}

The most consequential class was conflicting labels: 49 entries in
35 files gave 2,602 frames two different labels, and in
15 of these entries an identical frame range had been entered under two views
(Figure~\ref{fig:figure3_expert_errors}c). A further 15 ranges in
14 files extended beyond the end of their file, labeling 1,511
frames beyond the frame count read by CAT. Gaps left 5,930 frames unlabeled in 60 files,
and eight file rows carried no label at all. CAT prevents these errors by design, so its zero error count is not an empirical estimate (Table~\ref{tab:table3_error_prevention}).%

Errors were more frequent in CHD studies, affecting 72 of 659
files (10.9\%) compared with 37 of 1,087 normal files (3.4\%).
The mean per-study error rate was 7.0 percentage points higher
in CHD studies (permutation \textit{P} = 0.0012).
In the GEE model, CHD status (odds ratio 1.90, 95\% CI 1.13--3.18)
and longer files (odds ratio 1.93 per unit of log frame count,
95\% CI 1.31--2.82) were associated with higher error odds.
Differences in case complexity and annotators may also
contribute to the observed association with CHD.

\subsection{Agreement between traditional and CAT annotation}

On frames with a single spreadsheet-assigned view, agreement
between methods was 78.5\% (95\% CI 75.4--81.7), with
$\kappa=0.64$ and AC1 = 0.76. Agreement was higher in normal
than CHD studies (86.3\% vs 72.9\%) (Figure~\ref{fig:figure4_expert_agreement}a).
Agreement was lowest for frames CAT labeled as RVOT, 3VV,
or 3VT, which spreadsheets often assigned to NT or an
adjacent view. Across all traditional frames, 74.9\% agreed
with CAT, 21.2\% differed, and 3.9\% were unlabeled or
carried conflicting labels, making them unsuitable as
single training labels
(Figure~\ref{fig:figure4_expert_agreement}b).

CAT recorded more view segments per clip than spreadsheets
(3.6 vs 1.5). Within a tolerance of five frames, 52.9\%
of spreadsheet view-change boundaries matched a CAT boundary,
compared with only 13.4\% in the reverse direction
(Figure~\ref{fig:figure4_expert_agreement}c).
For frames assigned the same target view, image-quality
scores agreed on 60.4\% (weighted $\kappa=0.31$,
95\% CI 0.21--0.36).

\DefineFigure{figure4_expert_agreement}{wide}
{%
  \FloatGraphic{\includegraphics{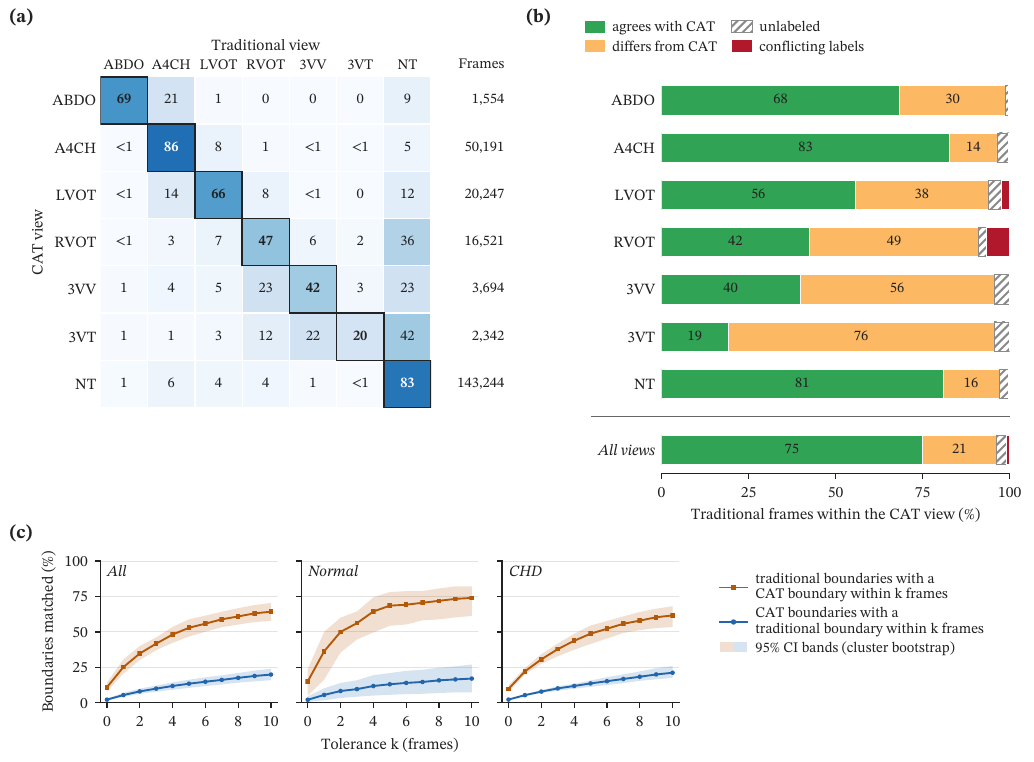}}%
}
{%
  Agreement between traditional and CAT
  annotation of the same files in the expert audit (1,746 files,
  25 studies). %
  \textbf{(a)}~View confusion on frames with exactly one traditional view; each
  row (CAT view) sums to 100\%. $<$1 denotes nonzero but under 1\%. Overall agreement
  78.5\% (95\% CI 75.4--81.7); Cohen's $\kappa$ 0.64 (CI
  0.60--0.68).
  \textbf{(b)}~Label status of all 247,477 traditional frames, by CAT view
  (percentages of the row).
  \textbf{(c)}~Share of view-change boundaries of one method with a boundary of
  the other within $\pm k$ frames, for all studies and separately for normal
  and CHD studies.
  View abbreviations as in
  Figure~\ref{fig:figure2_cat_software}.%
}
\placefigure{figure4_expert_agreement}

\subsection{Student annotation efficiency}
Median annotation time was 102.5 min with CAT versus 174.5 min
with spreadsheets, with five of six students completing annotations
faster using CAT (\textit{P} = 0.063). Labeled frames per minute
(ratio 1.77, 96.9\% CI 1.33--2.43) and correctly labeled frames
per minute (ratio 1.68, 1.25--2.36) were higher with CAT for all
six students (both \textit{P} = 0.031). These measures included
frames covered by CAT's default and block labels.
Reference segments per minute were higher for five students
(ratio 1.60, \textit{P} = 0.094)
(Table~\ref{tab:table4_student_outcomes} and
Figure~\ref{fig:figure5_student_crossover}a).

\DefineTable{table4_student_outcomes}{wide}
{%
  \begin{minipage}{\textwidth}
  \fontsize{8}{10}\selectfont
  \setlength{\tabcolsep}{4pt}%
  \begin{tabularx}{\textwidth}{@{}>{\raggedright\arraybackslash}X
                                  >{\raggedleft\arraybackslash}p{22mm}
                                  >{\raggedleft\arraybackslash}p{22mm}
                                  >{\raggedleft\arraybackslash}p{29mm}
                                  >{\centering\arraybackslash}p{15mm}
                                  >{\raggedleft\arraybackslash}p{9mm}@{}}
    \toprule
    \textbf{Measure}
      & \textbf{Traditional, median (range)}
      & \textbf{CAT, median (range)}
      & \textbf{Effect (96.9\% CI)}
      & \textbf{Students favoring CAT}
      & \textbf{\textit{P}} \\
    \midrule
    Annotation time, min
      & 174.5 (87--216) & 102.5 (54--219)
      & 0.62 (0.39--1.17) & 5/6\textsuperscript{\textdagger} & 0.063 \\
    Reference segments/min
      & 2.33 (1.19--2.94) & 3.29 (1.85--7.52)
      & 1.60 (0.96--3.57) & 5/6 & 0.094 \\
    Labeled frames/min
      & 34.9 (19.0--69.1) & 60.2 (25.3--116.1)
      & 1.77 (1.33--2.43) & 6/6 & 0.031 \\
    Correct frames/min
      & 19.4 (11.9--44.2) & 36.2 (14.9--67.5)
      & 1.68 (1.25--2.36) & 6/6 & 0.031 \\
    View accuracy, \%
      & 58.3 (36.9--64.0) & 59.0 (42.1--61.5)
      & +1.0 ($-$5.8 to +15.1) & 3/6 & 0.719 \\
    \bottomrule
  \end{tabularx}
  \begin{tablenotes}
    Effect: Hodges--Lehmann estimate of the paired CAT vs traditional effect with 96.9\% CI, a ratio for annotation time and the per-minute measures and a difference for view accuracy. \textit{P}: exact two-sided sign-flip test. \textsuperscript{\textdagger}Took less time with CAT.
  \end{tablenotes}
  \end{minipage}%
}
{%
  Annotation efficiency and view accuracy of the six medical students
  with traditional annotation and with CAT, against the expert reference%
}
\placetable{table4_student_outcomes}
The randomized crossover analysis yielded similar throughput
ratios of 1.63, 1.77, and 1.71 for reference segments, labeled
frames, and correctly labeled frames per minute, respectively
(all \textit{P} = 0.10). Excluding the student with uncertain
timing yielded corresponding ratios of 1.60, 1.88, and 1.87.
Analyses using CAT active time derived from event logs also
favored CAT, although comparable logs were unavailable
for spreadsheets.

Student spreadsheets contained 27 data-entry findings, including
10 provable major or critical errors. CAT output contained no
corresponding structural errors, as expected from its design
(Table~\ref{tab:table3_error_prevention}).

\DefineFigure{figure5_student_crossover}{wide}
{%
  \FloatGraphic{\includegraphics{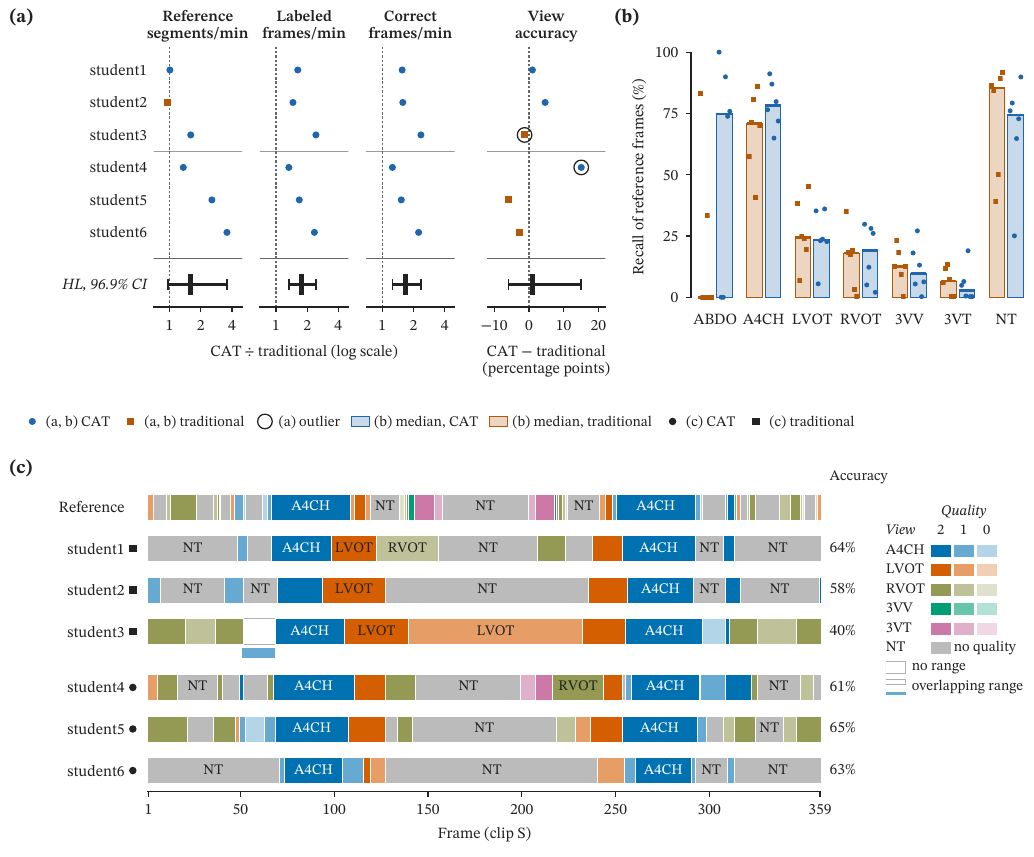}}%
}
{%
  Student crossover: six medical students each annotated one normal scan
  with each method (Group A, student1--student3, traditional first; Group B,
  student4--student6, CAT first), scored against the expert reference.
  \textbf{(a)}~Paired change per student: reference segments, labeled frames and correct
  frames per minute, as CAT\,$\div$\,traditional; view accuracy,
  as CAT\,$-$\,traditional. Below: Hodges--Lehmann (HL) estimate with exact
  96.9\% CI. Open rings mark outlying view accuracy values (robust $|z|$ above
  3.5).
  \textbf{(b)}~Recall of reference frames by view; bars, medians.
  \textbf{(c)}~Clip~S: the expert reference and each student's labels, with
  per-student view accuracy on the clip (as in a). Shade gives image quality; NT has none.
  View and quality abbreviations as in Figure~\ref{fig:figure2_cat_software}.%
}
\placefigure{figure5_student_crossover}

\subsection{Student view accuracy}
Median view accuracy was 58.3\% with spreadsheets and 59.0\%
with CAT, with no detectable difference between methods
(paired difference +1.0 percentage point, 96.9\% CI
$-$5.8 to +15.1, \textit{P} = 0.719)
(Table~\ref{tab:table4_student_outcomes}).
Three of six students achieved higher accuracy with CAT
(Figure~\ref{fig:figure5_student_crossover}a).
With both methods, recall was lowest for outflow-tract
and vessel views (Figure~\ref{fig:figure5_student_crossover}b,c).

\section{Discussion}\label{sec:discussion}

\subsection{Principal findings}

The expert audit found provable major or critical data-entry
errors in 6.2\% of spreadsheet-annotated files, more frequently
in CHD studies. CAT prevents these structural errors by design.
CAT annotations also contained more view segments per clip,
although this does not establish more accurate boundaries.
Agreement on image-quality grades was low, suggesting the need
for clearer scoring guidelines and annotator calibration.
In medical students, frame-based annotation throughput was
higher with CAT, without a detectable difference in view accuracy.

\subsection{Sources of spreadsheet errors}

Spreadsheet annotation requires manually transferring frame %
ranges and labels from a separate cine viewer. Without checks
for file length, overlapping ranges, or missing frames, invalid
and contradictory entries may go unnoticed. Even with restricted
label choices, student spreadsheets contained 10 provable major
or critical errors. CAT avoids manual frame-range entry by using %
fixed labeling keys and deriving ranges from annotation change
points, preventing these structural errors
(Table~\ref{tab:table3_error_prevention}). However, it cannot
prevent clinically incorrect but otherwise valid labels.

\subsection{Errors in CHD studies}

The higher error rate in CHD studies is concerning because
abnormal cases are relatively uncommon in screening datasets
and are important for training diagnostic models~\cite{Vega2025CARDIUM}.
This association may partly reflect
differences in file length, case complexity, or annotators.
Our error detectors identify logical and structural
inconsistencies but cannot detect incorrect labels that remain
internally consistent. Switching between a cine viewer and a
separate spreadsheet may introduce additional recording errors,
although this was not directly assessed. These findings
highlight the importance of validating spreadsheet annotations
before model training.

\subsection{Throughput and accuracy in students}

All three throughput measures favored CAT. Labeled and
correctly labeled frames per minute were higher for all six
students, while reference segments per minute was higher
for five of six students but not significantly
(\textit{P} = 0.094). The frame-based measures include
automatically assigned default labels, whereas reference
segments per minute is less affected by labeling style.
The crossover analysis yielded similar effect estimates,
although these did not reach statistical significance.
View accuracy did not differ detectably between methods.
Recall was lowest for outflow-tract and vessel views with
both methods (Figure~\ref{fig:figure5_student_crossover}b),
suggesting that recognizing these views remains challenging
regardless of the annotation method.

\subsection{Comparison with existing tools}

Existing tools support ultrasound video annotation but differ
in their scope and capabilities. The Oxford fetal heart
annotation tool supports frame-level cardiac view and
visibility labeling in AVI videos~\cite{Bridge2021HeartAnnotation}.
Annotation Web supports image-quality assessment alongside
other annotation tasks in ultrasound videos~\cite{Smistad2021AnnotationWeb}.
CAT combines direct DICOM
cine review with frame-level cardiac view and image-quality
labeling, using predefined labels and automatic frame-range
generation to reduce manual data entry. %

\subsection{Limitations}

The expert audit lacked an independent reference standard,
so disagreements between methods could not be adjudicated.
Spreadsheet and CAT annotations were performed in a fixed
order, 3--5 months apart, by partly different experts.
Consequently, differences may reflect both the annotation
method and the annotators. Expert annotation time was not
recorded, so potential time savings for cardiologists remain
unmeasured.

The student study included only six medical students who were
aware of the annotation method and used simplified scans
rather than DICOM files. Scan and period coincided within
each crossover group, limiting their separation in the
paired analysis. Annotation time was self-reported, with
event-log verification available only for CAT. The expert
reference was prepared using CAT, sharing its annotation
granularity. Finally, CAT was developed and evaluated by
the same research group.

\section{Conclusion}\label{sec:conclusion}
CAT streamlines frame-level annotation of fetal echocardiography,
allowing fetal cardiologists to focus on image interpretation rather
than manual spreadsheet entry. Our evaluations showed that CAT enabled %
finer-grained view annotations, avoided structural data-entry errors,
and improved annotation efficiency among medical students without a
detectable difference in view-label accuracy. These findings support
CAT as a practical tool for preparing fetal echocardiography datasets
for AI research. Future work will focus on two extensions: integrating a machine-learning
module that predicts standard cardiac views and adapts to annotator
corrections, and developing a study-level dashboard to track annotated
views and identify remaining annotation gaps.

\bmsubsection*{Funding}
Research reported in this publication was supported by the National Institute of General
Medical Sciences of the National Institutes of Health under Award Number P20GM152326, the Child
Health Research Institute AI pilot, and the University of Nebraska Collaboration Initiative
Grant from the Nebraska Research Initiative (NRI). This project is also supported by the U.S.
National Science Foundation under Award Numbers 2500836 and 2614824, the National Cancer
Institute of the National Institutes of Health under Award Number R03CA317707, and the National
Institute on Alcohol Abuse and Alcoholism of the National Institutes of Health under Award
Number R21AA032098. This research was supported by the State of Nebraska through the Pediatric
Cancer Research Group, part of the Child Health Research Institute. The content is solely the
responsibility of the authors and does not necessarily represent the official views of the
funding organizations.

\bmsubsection*{Ethics Statement}
This study was conducted under an Institutional Review Board (IRB)-approved protocol (\#0342-24-EP) and in accordance with the principles of the Declaration of Helsinki. The approved protocol encompassed both the retrospective use of fetal ultrasound images and image annotation by clinical experts and medical students as research personnel. Retrospective fetal ultrasound images were de-identified and assigned unique study identification numbers. To ensure confidentiality, annotators were identified using pseudonyms (Expert 1–4 and Student 1–6), and patient identifiers were excluded from all annotation datasets and analytical outputs.

\bmsubsection*{Consent}
The Institutional Review Board waived the requirement for informed consent for the analysis of
de-identified imaging data.

\bmsubsection*{Conflicts of Interest}
The authors declare no conflicts of interest.

\bmsubsection*{Data Availability Statement}
The fetal echocardiography studies contain protected health information and cannot be shared.
CAT is available from the corresponding author for non-commercial research use on reasonable request.

\printreferences

\printfloats
\end{document}